\documentclass[preprint,nofootinbib,prd,aps]{revtex4}
\usepackage{graphicx,amstext,amssymb,amsmath,color}
\begin{document}

\title{Quantum local-equilibrium  state with fixed multiplicity constraint and Bose-Einstein momentum correlations}

\author{M.D. Adzhymambetov$^{1}$}
\author{S.V. Akkelin$^{1}$}
\author{Yu.M. Sinyukov$^{1}$${^,}$${^2}$}
\affiliation{$^1$Bogolyubov Institute for Theoretical Physics,
Metrolohichna  14b, 03143 Kyiv,  Ukraine,\\
$^2$Warsaw University of Technology, Faculty of Physics
Koszykowa 75, 00-662 Warsaw, Poland }

\begin{abstract}

The one- and two-boson momentum spectra  are derived in the quantum local-equilibrium  canonical
ensemble of noninteracting bosons  with a fixed particle number constraint. We define the canonical ensemble  as a
  subensemble of  events associated with the grand-canonical ensemble.
Applying simple hydro-inspired  parametrization with  parameter values that
correspond roughly to the values at the system's breakup in $p+p$ collisions at the LHC energies, 
we  compare our findings with the treatment which is based on the grand-canonical ensembles where 
 mean particle numbers coincide with fixed particle numbers in the canonical ensembles. 
 We observe a significantly greater sensitivity of the
two-particle momentum correlation functions to fixed multiplicity constraint  compared to one-particle 
momentum spectra. The results of our analysis may be useful for interpretation of multiplicity-dependent
measurements of $p+p$ collision events.

\end{abstract}


 \maketitle

\section{Introduction}

Inasmuch as mean particle multiplicities in relativistic heavy ion collisions are large,
the whole set of  collision events at a fixed energy of nuclear collisions is typically  divided into subsets with
fixed charged-particle multiplicities.
Corresponding multiplicity classes are associated with collision centralities and, 
thereby, with the initial  system’s geometry which is primarily
characterized by the overall shape of the interaction region. 
 This  makes it possible to study  the multiplicity dependence of various observables measured at the same 
 energy of collisions.  In particular, the fixed particle multiplicity technique has
been utilized for analysis of the Bose-Einstein momentum correlations of
identical particles. These correlations are typically 
represented in terms of the interferometry radii. They are the result of the Gaussian fit 
of the correlation function defined as a ratio of the two-particle spectra to the product
of the single-particle ones. These radii reflect the space-time structure and dynamical evolution
of the systems created in nuclear collisions (for review of the correlation femtoscopy method  see e.g.
Ref. \cite{Sin-1}). One notable feature of these
measurements is that the effective system's volume, when extracted
from the Gaussian interferometry  radii, appears to  scale nearly linearly with charged particle multiplicity
(see e.g. Ref. (\cite{Alice})). 
This observation is in agreement with the hydrodynamical picture of nuclear collisions.

Recently, because of  the start of LHC 
experiments, the fixed particle multiplicity technique has 
been utilized for analysis of the Bose-Einstein momentum correlations of
identical particles  in proton-proton collisions at a fixed energy of collisions.
It was observed, in particular,   that 
measured in these collisions interferometry correlation radius parameters do not increase
with multiplicity  at   high
charged-particle multiplicities  \cite{Atlas,CMS}. While an  explanation of this effect is still absent,
it is suggestive to assume  
that the saturation effect in the multiplicity dependence of the
interferometry correlation radius parameters takes place 
once  the maximal overlap of colliding nucleons  is achieved in most central collisions. 
Indeed, the color glass condensate  effective theory predicts that once  maximal overlap is achieved 
higher multiplicities  can only be reached  by certain color charge 
fluctuations, which do not increase the initial size
of the system \cite{Lerran}.
Then, one can speculate that an  individual system 
created in a high-multiplicity
$p+p$ collision can be regarded as an element of a quantum-statistical ensemble of systems with
various numbers of particles   produced under the same initial-state geometry. 

In a quantum-statistical framework, observables are the expectation values of the corresponding
quantum operators with respect to a suitable statistical operator. For example, successful applicability of
almost perfect relativistic hydrodynamics   for the  description of a particle production
in relativistic heavy ion collisions (for a recent review
see, e.g., Ref. \cite{Schen})  indicates that actual state of a system created in collisions with the same  centrality
can be approximated by a
local-equilibrium statistical  
operator $\rho^{\text{leq}}$,  $\rm{Tr}[\rho^{\text{leq}}]=1$, which is obtained by maximizing 
the von Neumann entropy, $S=-{\rm Tr} [\rho \ln{\rho}]$,  with constrained mean values of energy-momentum
and conserved charge densities
on a given three-dimensional hypersurface (see, e.g., Ref. \cite{Zubarev}).
It is noteworthy that  high-multiplicity proton-proton 
collisions exhibit   collective behavior similar to that observed
in relativistic nuclear collisions. It indicates  that a hydrodynamic description
of  matter formed in these collisions  might also be possible \cite{Schen}. Application of fixed
high-multiplicity constraint to 
$p+p$ collision events   means then selecting some subensemble of  events
with the same initial-state geometry to which
the considered system belongs.  To assign a quantum statistical state to a subensemble of 
events with fixed multiplicity,   one can
utilize the  projection operator ${\cal P}_{N}$,
which  automatically invokes such a  constraint.
The aim of this work (see also Ref. \cite{Akkelin-2}) is to clarify how imposed particle number constraint affects
the one-particle spectra  and two-boson
momentum correlations in a quantum-field  local-equilibrium  state.
It is worth noting that for fairly high   particle numbers 
 a canonical  ground-state Bose-Einstein condensation  can occur. 
Such a condensation could, in principle, lead  to noticeable effects in  particle momentum spectra and 
correlations at fixed multiplicities. This issue is, however,
beyond the scope of this paper.\footnote{For such an
analysis, the   ground-state of the local-equilibrium statistical operator should be specified, 
and canonical  Bose condensation
in the corresponding ground state should  be taken into account.  For simple nonrelativistic quantum-field  models, 
it was done in Ref. \cite{Akkelin-1}, in which the relations of the ground-state Bose-Einstein condensation
at a fixed particle
number constraint to the particle  momentum spectra and correlations were  discussed.} 

\section{Local-equilibrium  statistical operator}

As a  starting point, we consider the quasiequilibrium state (see, e.g., Ref. \cite{Zubarev}) of  a real relativistic scalar
field. This state is represented by the  statistical operator $\rho^{\text{q}}(\sigma)$ as (we use the convention
$g^{\mu \nu}=\mbox{diag}(+1,-1,-1,-1)$)
\begin{eqnarray}
\rho^{\text{q}}(\sigma) =\frac{1}{Z^{\text{q}}(\sigma)}\hat{\rho}^{\text{q}}(\sigma),
\label{1.1} \\
\hat{\rho}^{\text{q}}(\sigma) =\exp \left ( - \int_{\sigma} d \sigma n_{\mu} (x) \beta_{\nu} (x) T^{\mu \nu} (x)
\right ), \label{1.2} 
\end{eqnarray}
where $\sigma$ is a three-dimensional
spacelike hypersurface with a timelike normal vector $n_{\mu}(x)$; 
$\beta_{\nu}(x) = \beta (x) u_{\nu}(x)$,  $u_{\mu}(x)u^{\mu}(x)=1$  are the 
corresponding Lagrange multipliers ($\beta = 1/T$ is the inverse temperature, and $u_{\mu}$ is the  $4$-velocity) 
on the hypersurface $\sigma$, adjusted such as to satisfy 
the actual mean  values of energy and momentum density at this hypersurface;
$Z^{\text{q}}(\sigma)$
is the normalization factor making
$\rm{Tr}[\rho^{\text{q}}(\sigma)]=1$; and $T^{\mu \nu} (x)$ is a scalar-field energy-momentum  tensor. For simplicity,
we disregard field self-interactions and consider a noninteracting  scalar
quantum field model. Then, the  $T^{\mu \nu} (x)$ reads 
\begin{eqnarray}
T^{\mu \nu}(x)=\partial^{\mu}\phi\partial^{\nu}\phi -g^{\mu \nu}L,
\label{2}
\end{eqnarray}
where the Lagrangian density is
\begin{eqnarray}
L=\frac{1}{2} \left (\frac{\partial
\phi}{\partial t}\right )^{2} - \frac{1}{2} \left (\frac{\partial
\phi}{\partial \textbf{r}}\right )^{2} - \frac{m^{2}}{2}\phi^{2}.
\label{3}
\end{eqnarray}
Here 
\begin{eqnarray}
\phi(x)  =
\int\frac{d^{3}p}{\sqrt{2\omega_{p}}}\frac{1}{(2\pi)^{3/2}}\left
(e^{-i\omega_{p}t+i \textbf{p} \textbf{r}} a (\textbf{p})+
e^{i\omega_{p}t-i \textbf{p} \textbf{r}} a^{\dag}(\textbf{p})\right
), \label{4}
\end{eqnarray}
and
\begin{eqnarray}
\omega_{p} =  \sqrt{\textbf{p}^{2} + m^{2}}.
\label{5}
\end{eqnarray}
 The quantization
prescription 
means that   $a^{\dag}(\textbf{p})$ and $a(\textbf{p})$
are creation and annihilation operators, respectively, which
satisfy the following canonical commutation relations:
\begin{eqnarray}
[a(\textbf{p}), a^{\dag}(\textbf{p}')] =
\delta^{(3)}(\textbf{p}-\textbf{p}') \label{6}
\end{eqnarray}
and 
$[a(\textbf{p}), a(\textbf{p}')]=[a^{\dag}(\textbf{p}), a^{\dag}(\textbf{p}')]=0$.

Before proceeding  further, let us digress for a moment and  consider the simple  case 
of the covariant global-equilibrium state, where the $\beta_{\mu}$ does not depend on spacetime coordinates 
across the infinite three-dimensional hypersurface. Then the statistical operator reads
\begin{eqnarray}
\rho^{\text{eq}} =\frac{1}{Z^{\text{eq}}}\exp \left ( -  \beta_{\mu}  P^{\mu} \right ),
\label{8} 
\end{eqnarray}
where $P^{\mu}=\int_{t} d^{3}r   T^{ \mu 0 } (x)$ is  $4$-momentum of the field defined at $t=\text{const}$ 
hypersurface. Then, using Eqs. (\ref{2}), (\ref{3}), and (\ref{4}),
we obtain 
\begin{eqnarray}
P^{\mu} = \frac{1}{2}\int d^{3}k k^{\mu}(a^{\dag}(\textbf{k})a(\textbf{k} ) + a(\textbf{k})a^{\dag}(\textbf{k} )).
\label{8.01}
\end{eqnarray}
It is convenient to introduce 
\begin{eqnarray}
P^{\mu}_{\text{reg}} = P^{\mu}- \langle 0 | P^{\mu} |0 \rangle =\int d^{3}k k^{\mu}a^{\dag}(\textbf{k})a(\textbf{k} ) ,
\label{8.2}
\end{eqnarray}
where $|0 \rangle$ is the   quantum field vacuum state, $a(\textbf{p})|0 \rangle = 0$.
Then, Eq. (\ref{8})  can be rewritten  as  
\begin{eqnarray}
\rho^{\text{eq}} =\frac{1}{Z^{\text{eq}}_{\text{reg}}}\exp \left ( -  \beta_{\mu}  P^{\mu}_{\text{reg}} \right ).
\label{8.1} 
\end{eqnarray}

It can be shown, e.g., by  Gaudin's method \cite{Gaudin}, that the 
statistical operator (\ref{8.1}) is  associated with the homogeneous 
ideal gas Bose distribution,
\begin{eqnarray}
f_{\text{eq}}(p)= \frac{1}{(2\pi)^{3}}\frac{1}{e^{\beta_{\nu}p^{\nu}}-1}.
\label{9}
\end{eqnarray}
Below, for the reader’s convenience, we
present an elementary derivation of it (see also Ref. \cite{Groot}). Let us start by defining 
 $a(\textbf{p},\alpha)$, 
\begin{eqnarray}
a(\textbf{p},\alpha) =  \exp \left ( \alpha \beta_{\mu} P^{\mu}_{\text{reg}}\right ) a(\textbf{p}) \exp \left ( - \alpha \beta_{\mu} P^{\mu}_{\text{reg}} \right ).
\label{10} 
\end{eqnarray}
Note that $a(\textbf{p}, 0) = a(\textbf{p}) $. 
Expression  (\ref{10}) implies that  $a(\textbf{p},\alpha)$ 
satisfies  equation 
\begin{eqnarray}
\frac{\partial a(\textbf{p},\alpha)}{\partial \alpha} = \left [     \beta_{\mu} P^{\mu}_{\text{reg}}, a(\textbf{p},\alpha) \right ].
\label{11} 
\end{eqnarray}
Taking into account that 
\begin{eqnarray}
\left [ \beta_{\mu} P^{\mu}_{\text{reg}}, a(\textbf{p},\alpha) \right ] =  \exp \left ( \alpha \beta_{\mu} P^{\mu}_{\text{reg}}\right ) \left [   \beta_{\mu} P^{\mu}_{\text{reg}}, a(\textbf{p}) \right ] \exp \left ( - \alpha \beta_{\mu} P^{\mu}_{\text{reg}}\right ),
\label{11.1} 
\end{eqnarray}
this  yields then 
\begin{eqnarray}
\frac{\partial a(\textbf{p},\alpha)}{\partial \alpha} =  -  \beta_{\mu}p^{\mu} a(\textbf{p},\alpha).
\label{11.2} 
\end{eqnarray}
The solution of this equation is 
\begin{eqnarray}
 a(\textbf{p},\alpha) =  a(\textbf{p}) \exp{(- \alpha \beta_{\mu}p^{\mu})}.
\label{12} 
\end{eqnarray}

Our next step is to combine  the cyclic invariance of the trace, $\rm{Tr} [\rho^{\text{eq}} (\sigma)  a^{\dag}(\textbf{p}_{1})a(\textbf{p}_{2}) ] $,  and  Eqs. (\ref{10}) and  (\ref{12}). Using  the cyclic invariance of the trace
and Eq. (\ref{10}), we obtain 
\begin{eqnarray}
\rm{Tr} [\rho^{\text{eq}} (\sigma)  a^{\dag}(\textbf{p}_{1})a(\textbf{p}_{2}) ] = \rm{Tr} [a(\textbf{p}_{2}) \rho^{\text{eq}} (\sigma)  a^{\dag}(\textbf{p}_{1}) ] = \nonumber \\ \rm{Tr} [\rho^{\text{eq}} (\sigma) a(\textbf{p}_{2}, 1) a^{\dag}(\textbf{p}_{1}) ].
\label{13} 
\end{eqnarray}
Taking into account  Eqs. (\ref{6}) 
and  (\ref{12}),   the r.h.s. of the above equation can be rewritten as 
\begin{eqnarray}
 \rm{Tr} [\rho^{\text{eq}}(\sigma) a(\textbf{p}_{2}, 1) a^{\dag}(\textbf{p}_{1}) ]= \rm{Tr} [\rho^{\text{eq}}(\sigma)  a^{\dag}(\textbf{p}_{1})a(\textbf{p}_{2},1) ] + [a(\textbf{p}_{2},1),a^{\dag}(\textbf{p}_{1})] = \nonumber \\
  e^{- \beta_{\mu}p^{\mu}_{2}} \left( \rm{Tr} [\rho^{\text{eq}}(\sigma)  a^{\dag}(\textbf{p}_{1})a(\textbf{p}_{2}) ] + \delta^{(3)}(\textbf{p}_{2}-\textbf{p}_{1}) \right ).
\label{14} 
\end{eqnarray}
Substituting this into Eq. (\ref{13}) we finally have 
\begin{eqnarray}
\rm{Tr} [\rho^{\text{eq}} (\sigma)  a^{\dag}(\textbf{p}_{1})a(\textbf{p}_{2}) ]  = \delta^{(3)}(\textbf{p}_{1}-\textbf{p}_{2}) \frac{1}{e^{\beta_{\nu}(p_{1}^{\nu}+p_{2}^{\nu})/2}-1}.
\label{15} 
\end{eqnarray}
Utilization of  the  Fourier transformation of Eq. (\ref{15}) with respect to $\Delta \textbf{p} = \textbf{p}_{2}-\textbf{p}_{1}$ immediately results in the ideal gas Bose distribution function (\ref{9}).

Now,  going back to the quasiequilibrium statistical operator (\ref{1.1}), (\ref{1.2}), 
we  suppose   that 
$\beta (x)$ and $u_{\mu}(x)$ are slowly varying  functions 
across the three-dimensional  hypersurface $\sigma$. This makes it  possible to 
apply a local thermal equilibrium approximation (see, e.g.,  Refs. \cite{Zubarev,Fl}) 
of the statistical operator (\ref{1.1}), (\ref{1.2}). The local thermal equilibrium
is an approximate concept  which is usually associated with the possibility of defining a fluid cell,
i.e., with the  existence of a scale at which  the system
appears to be at homogeneous equilibrium. Therefore, this scale should be much smaller than the distance over
which the $\beta_{\mu}(x)=\beta (x)u_{\mu}(x)$ varies essentially. On the other hand,  this scale has
to be assumed  large enough from a microscopic point of view,  meaning that the
typical  microscopic correlation lengths   are much smaller than the  size of a cell. 

To avoid additional complications and formulate the idea more concretely, we restrict ourselves to the case when the timelike normal vector $n_{\mu}(x)$ of the hypersurface $\sigma$ coincides with the $4$-velocity field $u_{\mu}(x)$,
\begin{eqnarray}
n_{\mu}(x)=u_{\mu}(x).
\label{15.1} 
\end{eqnarray}
Then, we replace the integral in Eq. (\ref{1.2})  by the sum as 
\begin{eqnarray}
\int_{\sigma} d \sigma n_{\mu} (x) \beta_{\nu} (x) T^{\mu \nu} (x)) \approx \sum_{s}\beta_{\nu}(x_{s}) P^{\nu}(\sigma_{s}),
\label{16}
\end{eqnarray}
where 
\begin{eqnarray}
 P^{\nu}(\sigma_{s}) = \int_{\sigma_{s}}d \sigma_{\mu}  T^{\mu \nu} (x) \approx u_{\mu} (x_{s}) \int_{\sigma_{s}}d \sigma   T^{\mu \nu} (x),
\label{16.1}
\end{eqnarray} 
and  the integral in the above equation is taken  over the homogeneity region of the  $\beta_{\nu} (x)$ 
around some point  $x_{s}^{\mu}$.  The homogeneity region  is defined as
a region of  the three-dimensional  hypersurface $\sigma$ where $\beta_{\nu} (x)$ 
does not vary in a noticeable way. It is instructive to rewrite $\beta_{\nu}(x_{s}) P^{\nu}(\sigma_{s})$
in the comoving coordinate system where $\tilde{u}_{\mu} (\tilde{x}_{s})=(1,\textbf{0})$.  Then,
\begin{eqnarray}
\beta_{\nu}(x_{s}) P^{\nu}(\sigma_{s}) = \tilde{\beta}_{0} (\tilde{x}_{s}) \tilde{P}^{0}(\tilde{\sigma}_{s}), \label{16.1-1} \\
\tilde{P}^{0}(\tilde{\sigma}_{s}) = \int_{\tilde{t}_{s}}d^{3} \tilde{r} T^{0 0} (\tilde{x}),
\label{16.1-2}\\
\tilde{\beta}_{0} (\tilde{x}_{s}) = \beta (x_{s}),
\end{eqnarray}
and   $\tilde{t}_{s}=\text{const}$. The key assumption underlying the local-equilibrium approximation 
is that  characteristic size, $\tilde{L}$, of the corresponding volume element  
 is large enough, i.e., $\tilde{L}\gg 1/m$. This assumption has important consequences. In particular,
 by using Eqs. (\ref{2}),  (\ref{3}), and  (\ref{4}), one 
can show that contributions of $aa$ and $a^{\dag}a^{\dag}$ terms to the $\tilde{P}^{0}(\tilde{\sigma}_{s})$ can be
neglected. In a sense, this provides the local thermal equilibrium in the region $s$ around
$x_{s}^{\mu}$.\footnote{Then, in particular,
an ideal fluid approximation  with a corresponding form
of the energy-momentum tensor is approximately
valid; see, e.g., Ref. \cite{Zubarev}. For quasiequilibrium states 
characterized by strong $\beta_{\mu}(x)$ gradients,  corrections to local
thermal equilibrium approximation and, thereby, to  ideal fluid
approximation  are  sizeable and need to be taken into account.}  The corresponding local-equilibrium
statistical operator is 
\begin{eqnarray}
\rho^{\text{leq}}(\sigma) =\frac{1}{Z^{\text{leq}}_{\text{reg}}(\sigma)}\hat{\rho}^{\text{leq}}(\sigma),
\label{16.2} \\
\hat{\rho}^{\text{leq}}(\sigma) =\exp \left ( - \sum_{s}\beta_{\nu}(\sigma_{s}) P^{\nu}_{\text{reg}}(\sigma_{s})
\right ), \label{16.3} 
\end{eqnarray}
where $P^{\mu}_{\text{reg}} (\sigma_{s})= P^{\mu}(\sigma_{s})- \langle 0 | P^{\mu} (\sigma_{s}) |0 \rangle$.
By using Eqs. (\ref{2}),  (\ref{3}), and  (\ref{4}),   we  get 
\begin{eqnarray}
 \beta_{\nu}(\sigma_{s})P^{\nu}_{\text{reg}} (\sigma_{s})  \approx \nonumber \\ \beta(\sigma_{s}) \int \frac{d^{3}k}{k_{0}} \frac{d^{3}k'}{k'_{0}} 
 \frac{u_{\mu}(x_{s})k^{\mu}u_{\nu}(x_{s})k'^{\nu}}{(2\pi)^{3}}   \int_{\sigma_{s}}d \sigma  e^{i(k-k')x}  \sqrt{k_{0}k'_{0}}
   a^{\dag}(\textbf{k})a(\textbf{k}') .
\label{16.3-1}
\end{eqnarray} 
Going to the local rest frame for a cell, we can write Eq. (\ref{16.3-1})  in the
following form: 
\begin{eqnarray}
  \beta (x_{s}) \tilde{P}^{0}_{\text{reg}}(\tilde{\sigma}_{s})  \approx   \beta(\sigma_{s}) \int d^{3}\tilde{k}d^{3}\tilde{k}' \frac{1}{(2\pi)^{3}} \int_{\tilde{t}_{s}}d^{3}\tilde{r} e^{i(\tilde{k}-\tilde{k}')\tilde{x}}\sqrt{\tilde{k}_{0}\tilde{k}'_{0} } a^{\dag}(\tilde{\textbf{k}})a(\tilde{\textbf{k}}').
  \label{16.3-2}
\end{eqnarray}
Equation (\ref{16.3-1}) makes possible to rewrite the operator  $\sum_{s}\beta_{\nu}(\sigma_{s}) P^{\nu}_{\text{reg}}(\sigma_{s})$ as 
\begin{eqnarray}
\sum_{s}\beta_{\nu}(\sigma_{s}) P^{\nu}_{\text{reg}}(\sigma_{s}) \approx \int d^{3}k d^{3}k'    A (\textbf{k}, \textbf{k}',\sigma)a^{\dag}(\textbf{k})a(\textbf{k}') , \label{16.3-3} 
\end{eqnarray}
where 
\begin{eqnarray}
   A (\textbf{k}, \textbf{k}',\sigma) =\sum_{s} A_{s} (\textbf{k}, \textbf{k}',\sigma), \label{16.3-4} \\
   A_{s} (\textbf{k}, \textbf{k}',\sigma)=\beta(\sigma_{s})  \frac{1}{\sqrt{k_{0}k'_{0}}} 
 \frac{u_{\mu}(x_{s})k^{\mu}u_{\nu}(x_{s})k'^{\nu}}{(2\pi)^{3}}   \int_{\sigma_{s}}d \sigma  e^{i(k-k')x}  
  . \label{16.3-5} 
\end{eqnarray}
 
\section{Quantum local-equilibrium grand-canonical  ensemble}

In this section, we calculate one-particle and  two-particle  momentum spectra in the  grand-canonical ensemble, which is described by the local-equilibrium 
statistical operator.  For this aim,  it is 
convenient to compute  fist 
$\langle a^{\dag}(\textbf{p}_{1})a(\textbf{p}_{2}) \rangle$,  where $\langle ...  \rangle =\rm{Tr} [... \rho^{\text{leq}}(\sigma)]  $. It can be done by adapting  the Gaudin's method
to our problem.
We start by defining   $a(\textbf{p},\alpha)$, $a(\textbf{p}, 0) = a(\textbf{p}) $,  $\rm{Im}(\alpha) = 0$,
 as 
\begin{eqnarray}
a(\textbf{p},\alpha) =  \exp \left ( \alpha  \sum_{s}\beta_{\nu}(\sigma_{s}) P^{\nu}_{\text{reg}}(\sigma_{s})\right ) a(\textbf{p}) \exp \left ( - \alpha \sum_{s}\beta_{\nu}(\sigma_{s}) P^{\nu}_{\text{reg}}(\sigma_{s}) \right ),
\label{17} 
\end{eqnarray}
where $\sum_{s}\beta_{\nu}(\sigma_{s}) P^{\nu}_{\text{reg}}(\sigma_{s})$ is defined by Eqs.
(\ref{16.3-3}), (\ref{16.3-4}),
and (\ref{16.3-5}).
Applying the operator identity 
\begin{eqnarray}
 e^{X}Ye^{-X} =Y + \left[ X, Y\right] + \frac{1}{2!}\left [X, \left[ X, Y \right]\right ] +\frac{1}{3!}\left[X, \left[X, \left[ X, Y\right]\right ] \right ] + ...,
\label{17.1} 
\end{eqnarray}
we can write the result as 
\begin{eqnarray}
 a(\textbf{p},\alpha) =  a(\textbf{p}) + (- \alpha) \int d^{3}k  A(\textbf{p}, \textbf{k},\sigma)a(\textbf{k}) + \nonumber \\
 \frac{(- \alpha)^{2}}{2!} \int d^{3}k_{1}d^{3}k A(\textbf{p}, \textbf{k}_{1},\sigma)A(\textbf{k}_{1}, \textbf{k},\sigma) a(\textbf{k}) + \nonumber \\ \frac{(- \alpha)^{3}}{3!} \int d^{3}k_{1}d^{3}k_{2}d^{3}k A(\textbf{p}, \textbf{k}_{1},\sigma)A(\textbf{k}_{1}, \textbf{k}_{2},\sigma)A(\textbf{k}_{2}, \textbf{k},\sigma) a(\textbf{k}) + ... .
\label{17.2} 
\end{eqnarray}
Taking into account the nonoverlapping of different cells and neglecting the surface  effect
on the boundaries  of neighboring cells, we get
\begin{eqnarray}
 a(\textbf{p},\alpha) \approx a(\textbf{p}) + (- \alpha) \sum_{s} \int d^{3}k  A_{s}(\textbf{p}, \textbf{k},\sigma)a(\textbf{k}) + \nonumber \\
 \frac{(- \alpha)^{2}}{2!} \sum_{s} \int d^{3}k_{1}d^{3}k A_{s}(\textbf{p}, \textbf{k}_{1},\sigma)A_{s}(\textbf{k}_{1}, \textbf{k},\sigma) a(\textbf{k}) + \nonumber \\ \frac{(- \alpha)^{3}}{3!} \sum_{s} \int d^{3}k_{1}d^{3}k_{2}d^{3}k A_{s}(\textbf{p}, \textbf{k}_{1},\sigma)A_{s}(\textbf{k}_{1}, \textbf{k}_{2},\sigma)A_{s}(\textbf{k}_{2}, \textbf{k},\sigma) a(\textbf{k}) + ... .
\label{17.2-1} 
\end{eqnarray}
Substituting Eq. (\ref{16.3-5}) into Eq. (\ref{17.2-1}) and going to the local rest frame for each cell,
  we can perform an approximate integration over the momenta assuming that $\tilde{L}\gg 1/m$, where $\tilde{L}$ is the characteristic 
 length scale  of a cell. The result written in the laboratory  coordinate system is 
\begin{eqnarray}
 a(\textbf{p},\alpha) \approx a(\textbf{p}) + (- \alpha) \sum_{s}\int d^{3}k   A_{s}(\textbf{p}, \textbf{k},\sigma)a(\textbf{k})  + \nonumber \\
 \frac{(- \alpha)^{2}}{2!} \sum_{s} \int d^{3}k \left( \frac{(p^{\nu}+ k^{\nu})}{2} \beta_{\nu}(\sigma_{s}) \right)  A_{s}(\textbf{p}, \textbf{k},\sigma) a(\textbf{k}) + \nonumber \\ \frac{(- \alpha)^{3}}{3!}\sum_{s}\int d^{3}k \left( \frac{(p^{\nu}+ k^{\nu})}{2} \beta_{\nu}(\sigma_{s}) \right)^{2} A_{s}(\textbf{p}, \textbf{k},\sigma) a(\textbf{k}) + ....
\label{17.2-2} 
\end{eqnarray}
Taking into account that  main contribution in the integral over $\textbf{k}$ is given by $\textbf{k} \approx \textbf{p}$, it is convenient  to substitute $ u_{\mu}(x_{s})p^{\mu}u_{\nu}(x_{s})k^{\nu}$ in the  $A_{s}(\textbf{p}, \textbf{k},\sigma)$ by $((p^{\nu}+ k^{\nu})u_{\nu}(x_{s})/2)^{2}$. The result is    
\begin{eqnarray}
 a(\textbf{p},\alpha) \approx a(\textbf{p}) + (- \alpha) \sum_{s}\int d^{3}k \left( \frac{(p^{\nu}+ k^{\nu})}{2} \beta_{\nu}(\sigma_{s}) \right) \delta^{(3)}_{s}(\textbf{p}-\textbf{k}) a(\textbf{k}) + \nonumber \\
 \frac{(- \alpha)^{2}}{2!} \sum_{s} \int d^{3}k \left( \frac{(p^{\nu}+ k^{\nu})}{2} \beta_{\nu}(\sigma_{s}) \right)^{2} \delta^{(3)}_{s}(\textbf{p}-\textbf{k}) a(\textbf{k}) + \nonumber \\ \frac{(- \alpha)^{3}}{3!}\sum_{s}\int d^{3}k \left( \frac{(p^{\nu}+ k^{\nu})}{2} \beta_{\nu}(\sigma_{s}) \right)^{3} \delta^{(3)}_{s}(\textbf{p}-\textbf{k}) a(\textbf{k}) + ...,
\label{17.2-3} 
\end{eqnarray}
where we introduced notation
\begin{eqnarray}
  \delta^{(3)}_{s}(\textbf{p}-\textbf{k}) = \frac{u_{\mu} (x_{s})}{(2\pi)^{3}}\int_{\sigma_{s}}d \sigma  \frac{(p^{\mu}+k^{\mu})}{2\sqrt{\omega_{p}\omega_{k}}}
 e^{i(p-k)x}.
\label{17.2-4}
\end{eqnarray}
Note that the first term in Eq. (\ref{17.2-3}) may be written  as 
\begin{eqnarray}
 a(\textbf{p}) =  \int d^{3}k  \delta^{(3)}(\textbf{p}-\textbf{k}) a(\textbf{k}) \approx \int d^{3}k \sum_{s} \delta^{(3)}_{s}(\textbf{p}-\textbf{k}) a(\textbf{k}) .
\label{17.5} 
\end{eqnarray}
Substituting  (\ref{17.5})  into  (\ref{17.2-3}), we obtain 
\begin{eqnarray}
     a(\textbf{p}, \alpha) \approx \sum_{n=0}^{\infty} \frac{ (-\alpha)^{n}}{n!}\sum_{s} \int d^{3}k   \left(  \frac{k^{\nu} + p^{\nu} }{2} \beta_{\nu}(\sigma_{s})  \right)^{n}  \delta^{(3)}_{s}(\textbf{p}-\textbf{k}) a(\textbf{k}).
     \label{17.6} 
\end{eqnarray}
It is convenient to introduce notation
\begin{eqnarray}
     G_{\alpha}^{*}(\textbf{p}, \textbf{k}, \sigma) =
     \sum_{s}  \exp \left( -\alpha  \beta_{\mu}(\sigma_{s}) \frac{k^{\mu} + p^{\mu} }{2}  \right)  \delta^{(3)}_{s}(\textbf{p}-\textbf{k}), \label{17.7} 
\end{eqnarray}
and rewrite Eq. (\ref{17.6}) in the form 
\begin{eqnarray}
     a(\textbf{p}, \alpha) \approx \int d^{3}k G_{\alpha}^{*}(\textbf{p}, \textbf{k}, \sigma) a(\textbf{k}). \label{17.8} 
\end{eqnarray}
From Eq. (\ref{17.7}), we have that 
\begin{eqnarray}
 G_{\alpha}^{*}(\textbf{p}, \textbf{k},\sigma) = G_{\alpha}(\textbf{k}, \textbf{p},\sigma)
\label{17.9} 
\end{eqnarray}
and that\footnote{Note here that for $G_{\alpha}^{*}$, which  make Eq. (\ref{17.8}) an exact equality,  Eq. (\ref{17.10}) also becomes an exact equality. }
\begin{eqnarray}
\int d^{3}k G_{\alpha_{1}}^{*}(\textbf{p}_{2}, \textbf{k},\sigma)G_{\alpha_{2}}^{*}(\textbf{k}, \textbf{p}_{1},\sigma) \approx  G_{\alpha_{1}+\alpha_{2}}^{*}(\textbf{p}_{2}, \textbf{p}_{1},\sigma).
\label{17.10} 
\end{eqnarray}

We can now employ 
the cyclic invariance of the trace   to get expression for 
$\langle a^{\dag}(\textbf{p}_{1})a(\textbf{p}_{2}) \rangle$. By using  Eq. (\ref{17}) and the
cyclic invariance of the trace,  one can write 
\begin{eqnarray}
Tr [ \rho^{\text{leq}}(\sigma)  a^{\dag}(\textbf{p}_{1})a(\textbf{p}_{2}) ] = Tr [\rho^{\text{leq}}(\sigma) a(\textbf{p}_{2}, 1) a^{\dag}(\textbf{p}_{1}) ] = \nonumber \\ Tr [\rho^{\text{leq}}(\sigma)  a^{\dag}(\textbf{p}_{1})a(\textbf{p}_{2},1) ] + [a(\textbf{p}_{2},1),a^{\dag}(\textbf{p}_{1})].
\label{26} 
\end{eqnarray}
Here, $a(\textbf{p}_{2},1)$ is given by Eq. (\ref{17.8}).
By using Eq. (\ref{17.8}),  we can write
\begin{eqnarray}
 [a(\textbf{p}_{2},1),a^{\dag}(\textbf{p}_{1})] = G_{1}^{*}(\textbf{p}_{2}, \textbf{p}_{1},\sigma).
\label{27} 
\end{eqnarray}
Then, Eq. (\ref{26}) becomes  
\begin{eqnarray}
 \langle a^{\dag}(\textbf{p}_{1}) a(\textbf{p}_{2}) \rangle  =  Tr [ \rho^{\text{leq}}(\sigma)  a^{\dag}(\textbf{p}_{1})a(\textbf{p}_{2}) ] = \nonumber \\
 \int d^{3}k  G_{1}^{*}(\textbf{p}_{2}, \textbf{k}, \sigma)  \langle a^{\dag}(\textbf{p}_{1}) a(\textbf{k}) \rangle + 
  G_{1}^{*}(\textbf{p}_{2}, \textbf{p}_{1},\sigma).
\label{28} 
\end{eqnarray}
This equation can be solved by iteration. The result is 
\begin{eqnarray}
 \langle a^{\dag}(\textbf{p}_{1}) a(\textbf{p}_{2}) \rangle  =  
 G_{1}^{*}(\textbf{p}_{2}, \textbf{p}_{1},\sigma) +  \int d^{3}k  G_{1}^{*}(\textbf{p}_{2}, \textbf{k},\sigma)  
 G_{1}^{*}(\textbf{k}, \textbf{p}_{1},\sigma) + ....
\label{28.1} 
\end{eqnarray}
Taking into account (\ref{17.10}) we get 
\begin{eqnarray}
 \langle a^{\dag}(\textbf{p}_{1}) a(\textbf{p}_{2}) \rangle  = \sum_{n=1}^{\infty} 
 G_{n}^{*}(\textbf{p}_{2}, \textbf{p}_{1},\sigma) ,
\label{29} 
\end{eqnarray}
where $G_{n}^{*}$   is given by Eq. (\ref{17.7}). 
Substituting $G_{n}^{*}$ into Eq. (\ref{29}), we  have 
\begin{eqnarray}
\langle a^{\dag}(\textbf{p}_{1}) a(\textbf{p}_{2}) \rangle    \approx  \sum_{s} \frac{1}{
\exp \left ( \frac{(p^{\nu}_{1}+ p^{\nu}_{2})}{2} \beta_{\nu}(\sigma_{s}) \right ) -1 }
  \delta^{(3)}_{s}(\textbf{p}_{1}-\textbf{p}_{2}).
\label{35} 
\end{eqnarray}

Our next step is to replace sums over cells with integral over the hypersurface $\sigma$. This leads to 
\begin{eqnarray}
G_{n} (\textbf{p}_{1}, \textbf{p}_{2}, \sigma)  \approx  \frac{1}{(2\pi)^{3}\sqrt{p^{0}_{1}p^{0}_{2} }} \int_{\sigma}  d \sigma_{\mu}p^{\mu}e^{-i(p_{1}-p_{2})x} e^{-n\beta_{\nu} (x)p^{\nu}} ,
\label{36.1} \\
 \langle a^{\dag}(\textbf{p}_{1}) a(\textbf{p}_{2}) \rangle  \approx \frac{1}{(2\pi)^{3}\sqrt{p^{0}_{1}p^{0}_{2} }}  \int_{\sigma}  d \sigma_{\mu}p^{\mu}e^{-i(p_{1}-p_{2})x}  \frac{1}{e^{\beta_{\nu} (x)p^{\nu}}-1},
\label{37} 
\end{eqnarray}
where $p^{\mu}=(p^{\mu}_{1}+p^{\mu}_{2})/2$.
One-particle momentum spectra then read
\begin{eqnarray}
p_{0}\frac{d^{3}\langle N \rangle}{d^{3}p} = p_{0}\langle a^{\dag}(\textbf{p}) a(\textbf{p}) \rangle  \approx   \int_{\sigma}  d \sigma_{\mu}p^{\mu}f^{\text{leq}}(x,p),
\label{37.1} 
\end{eqnarray}
where  $f^{\text{leq}}(x,p)$ is the grand-canonical  distribution function,
which has the familiar form of the local-equilibrium distribution function of the 
relativistic ideal gas of bosons,
\begin{eqnarray}
f^{\text{leq}}(x,p)= \frac{1}{(2\pi)^{3}} \frac{1}{e^{\beta_{\nu} (x)p^{\nu}}-1}.
\label{37.2} 
\end{eqnarray}
It is worth noting that our derivation can be readily  extended to the local-equilibrium grand-canonical ensemble with 
nonzero constant chemical potential, $\mu$, associated with mean number of particles. 
Then, $u_{\nu}(x)p^{\nu} \rightarrow u_{\nu}(x)p^{\nu} - \mu$.

Evidently,  our derivation is rather heuristic and nonrigorous. But, in our opinion, it is 
instructive and adds some 
insights into the consistency  of the  approximations 
needed to associate quasiequilibrium statistical operator $\rho^{\text{q}}$  with  the 
local-equilibrium ideal Bose  gas distribution $f_{\text{leq}}(x,p)$.

Proceeding in the same way as above, one can readily  derive an expression for the two-particle momentum 
spectra, 
\begin{eqnarray}
p^{0}_{1}p^{0}_{2} \frac{d^{6}\langle N(N-1)\rangle }{d^{3}p_{1}d^{3}p_{2}} = p^{0}_{1}p^{0}_{2}  \langle a^{\dag}(\textbf{p}_{1}) a^{\dag}(\textbf{p}_{2}) a(\textbf{p}_{1}) a(\textbf{p}_{2}) \rangle.
\label{38} 
\end{eqnarray}
We start by using  the cyclic invariance of the
trace. This leads to 
\begin{eqnarray}
 \langle a^{\dag}(\textbf{p}_{1}) a^{\dag}(\textbf{p}_{2}) a(\textbf{p}_{1}) a(\textbf{p}_{2}) \rangle  = \nonumber \\
  \langle a^{\dag}(\textbf{p}_{1}) a(\textbf{p}_{1}) \rangle G_{1}^{*}(\textbf{p}_{2}, \textbf{p}_{2},\sigma) +
  \langle a^{\dag}(\textbf{p}_{2}) a(\textbf{p}_{1}) \rangle G_{1}^{*}(\textbf{p}_{2}, \textbf{p}_{1},\sigma)+ \nonumber \\ \int d^{3}k \langle a^{\dag}(\textbf{p}_{1}) a^{\dag}(\textbf{p}_{2}) a(\textbf{p}_{1}) a(\textbf{k}) \rangle
  G_{1}^{*}(\textbf{p}_{2}, \textbf{k},\sigma). 
\label{42} 
\end{eqnarray}
The above equation is solved by iteration. We obtain  
\begin{eqnarray}
\langle a^{\dag}(\textbf{p}_{1}) a^{\dag}(\textbf{p}_{2}) a(\textbf{p}_{1}) a(\textbf{p}_{2}) \rangle  = \nonumber \\ \langle a^{\dag}(\textbf{p}_{1}) a(\textbf{p}_{1}) \rangle   \sum_{n=1}^{\infty} 
 G_{n}^{*}(\textbf{p}_{2}, \textbf{p}_{2},\sigma)  + \langle a^{\dag}(\textbf{p}_{2}) a(\textbf{p}_{1}) \rangle   \sum_{n=1}^{\infty} 
 G_{n}^{*}(\textbf{p}_{2}, \textbf{p}_{1},\sigma). 
\label{43} 
\end{eqnarray}
Using  Eq. (\ref{29}), we can write the result as 
\begin{eqnarray}
\langle a^{\dag}(\textbf{p}_{1}) a^{\dag}(\textbf{p}_{2}) a(\textbf{p}_{1}) a(\textbf{p}_{2}) \rangle  = \nonumber \\ \langle a^{\dag}(\textbf{p}_{1}) a(\textbf{p}_{1}) \rangle   \langle a^{\dag}(\textbf{p}_{2}) a(\textbf{p}_{2}) \rangle   + \langle a^{\dag}(\textbf{p}_{2}) a(\textbf{p}_{1}) \rangle   \langle a^{\dag}(\textbf{p}_{1}) a(\textbf{p}_{2}) \rangle ,
\label{44} 
\end{eqnarray}
where $\langle a^{\dag} a \rangle$ are given by Eq. (\ref{37}). Equation (\ref{44}) is the 
particular case of the thermal Wick’s theorem \cite{Wick}.

Results of this section, in particular Eq. (\ref{36.1}),  will  be used in the next 
section to evaluate
particle momentum spectra and correlations at a fixed particle number constraint. 

\section{Quantum local-equilibrium canonical ensemble with
fixed particle number constraint}

We begin this section  by defining the local-equilibrium canonical ensemble with a fixed particle number 
constraint  as a subensemble of the corresponding
grand-canonical ensemble. For this aim,
 we apply the constraint to the
 statistical operator  given by Eq. (\ref{16.3}). It implies
 utilization of   the projection operator ${\cal P}_{N}$,
\begin{eqnarray}
{\cal P}_{N} = \int d^{3}p_{1}... d^{3}p_{N} |p_{1},...,p_{N}\rangle
\langle p_{1},...,p_{N} | ,  \label{45} \\
|p_{1},...,p_{N}\rangle =\frac{1}{\sqrt{N!}} a^{\dag}(\textbf{p}_{1})...a^{\dag}(\textbf{p}_{N})| 0 \rangle ,
\label{45.1}
\end{eqnarray}
which  automatically invokes the corresponding constraint. 
Then, the local equilibrium  statistical operator with the constraint,
$\rho^{\text{leq}}_{N}(\sigma)$, is\footnote{Below, for brevity,  we omit subscripts and superscripts
$\text{leq}$ and  $\text{reg}$ whenever it is clear from the context.}
\begin{eqnarray}
\rho_{N} (\sigma)  = \frac{1}{Z_{N}(\sigma)} \hat{\rho}_{N}(\sigma) , \label{47} \\
\hat{\rho}_{N}(\sigma)  = {\cal P}_{N}\hat{\rho}(\sigma) {\cal P}_{N}, 
 \label{46.1} \\
 Z_{N}(\sigma)  =\rm{Tr}[\hat{\rho}_{N} (\sigma) ] , \label{46}
\end{eqnarray}
and we define  $\langle  ... \rangle_{N} = \rm{Tr}[\rho_{N} ....]$. To evaluate 
two-boson momentum spectra at a fixed multiplicity, $\langle a^{\dag}(\textbf{p}_{1})a^{\dag}(\textbf{p}_{2}) a(\textbf{p}_{1})a(\textbf{p}_{2}) \rangle_{N}$, we will follow 
the same strategy as in the previous section.
We begin with evaluation of  $\langle a^{\dag}(\textbf{p}_{1})a^{\dag}(\textbf{p}_{2})  \rangle_{N}$. This 
can be done by using its invariance under cyclic permutations.  One gets 
\begin{eqnarray}
\langle a^{\dag}(\textbf{p}_{1})a^{\dag}(\textbf{p}_{2})  \rangle_{N} = \rm{T}r[\rho_{N} (\sigma) a^{\dag}(\textbf{p}_{1}) a(\textbf{p}_{2})]= \nonumber \\ \rm{Tr}[a(\textbf{p}_{2}) \rho_{N} (\sigma) a^{\dag}(\textbf{p}_{1}) ] = \frac{1}{Z_{N}(\sigma)}Tr[a(\textbf{p}_{2}) {\cal P}_{N}\hat{\rho}(\sigma) {\cal P}_{N} a^{\dag}(\textbf{p}_{1})].
 \label{49}
\end{eqnarray}
Utilizing elementary operator algebra,
one can  prove that
\begin{eqnarray}
a(\textbf{p}_{2}) {\cal P}_{N} = {\cal P}_{N-1} a(\textbf{p}_{2}).
 \label{50}
\end{eqnarray}
We also have 
\begin{eqnarray}
a(\textbf{p}_{2}) \hat{\rho}(\sigma) = \hat{\rho}(\sigma)  a(\textbf{p}_{2},1),
 \label{51}
\end{eqnarray}
where $a(\textbf{p}_{2},1)$ and $\hat{\rho}(\sigma)$ are defined by Eqs. (\ref{17}) and (\ref{16.3}), respectively.
Therefore the  r.h.s. of Eq. (\ref{49}) can be rewritten as 
\begin{eqnarray}
\frac{1}{Z_{N}(\sigma)}\rm{Tr}[a(\textbf{p}_{2}) {\cal P}_{N}\hat{\rho}(\sigma) {\cal P}_{N} a^{\dag}(\textbf{p}_{1})] = \frac{1}{Z_{N}(\sigma)}\rm{Tr}[ {\cal P}_{N-1}\hat{\rho}(\sigma) a(\textbf{p}_{2},1){\cal P}_{N} a^{\dag}(\textbf{p}_{1})].
 \label{52}
\end{eqnarray}
Next, using Eqs. (\ref{17.8}) and  (\ref{50}), we obtain  $a(\textbf{p}_{2},1){\cal P}_{N} ={\cal P}_{N-1}  a(\textbf{p}_{2},1) $.
Therefore,
\begin{eqnarray}
\frac{1}{Z_{N}(\sigma)}\rm{Tr}[ {\cal P}_{N-1}\hat{\rho}(\sigma) a(\textbf{p}_{2},1){\cal P}_{N} a^{\dag}(\textbf{p}_{1})] = \frac{1}{Z_{N}(\sigma)}\rm{Tr}[ {\cal P}_{N-1}\hat{\rho}(\sigma) {\cal P}_{N-1} a(\textbf{p}_{2},1) a^{\dag}(\textbf{p}_{1})] = \nonumber \\
\frac{1}{Z_{N}(\sigma)}\rm{Tr}[ {\cal P}_{N-1}\hat{\rho}(\sigma) {\cal P}_{N-1} a^{\dag}(\textbf{p}_{1}) a(\textbf{p}_{2},1) ]+ \frac{1}{Z_{N}(\sigma)}[a(\textbf{p}_{2},1), a^{\dag}(\textbf{p}_{1})]Tr[ {\cal P}_{N-1}\hat{\rho}(\sigma) {\cal P}_{N-1}].
 \label{53}
\end{eqnarray}
Furthermore, accounting for Eq. (\ref{17.8})  one can see that 
\begin{eqnarray}
\frac{1}{Z_{N}(\sigma)}\rm{Tr}[ {\cal P}_{N-1}\hat{\rho}(\sigma) {\cal P}_{N-1} a^{\dag}(\textbf{p}_{1}) a(\textbf{p}_{2},1) ]=  \frac{Z_{N-1}(\sigma)}{Z_{N}(\sigma)} \rm{Tr}[\rho_{N-1} (\sigma) a^{\dag}(\textbf{p}_{1}) 
a(\textbf{p}_{2},1)] = \nonumber \\ \frac{Z_{N-1}(\sigma)}{Z_{N}(\sigma)} \langle a^{\dag}(\textbf{p}_{1}) a(\textbf{p}_{2},1) \rangle_{N-1} =\frac{Z_{N-1}(\sigma)}{Z_{N}(\sigma)}\int d^{3}k G_{1}^{*}(\textbf{p}_{2}, \textbf{k},\sigma)\langle a^{\dag}(\textbf{p}_{1}) a(\textbf{k}) \rangle_{N-1} 
 \label{53.1}
\end{eqnarray}
and that
\begin{eqnarray}
\frac{1}{Z_{N}(\sigma)}[a(\textbf{p}_{2},1), a^{\dag}(\textbf{p}_{1})]\rm{Tr}[ {\cal P}_{N-1}\hat{\rho}(\sigma) {\cal P}_{N-1}] = \frac{Z_{N-1}(\sigma)}{Z_{N}(\sigma)} [a(\textbf{p}_{2},1), a^{\dag}(\textbf{p}_{1})] = \nonumber \\
 \frac{Z_{N-1}(\sigma)}{Z_{N}(\sigma)} G_{1}^{*}(\textbf{p}_{2}, \textbf{p}_{1},\sigma).
 \label{53.2}
\end{eqnarray}
Substituting Eqs. (\ref{53.1}) and (\ref{53.2}) into Eq. (\ref{53}) and then into the  r.h.s. of Eq. (\ref{49}),
we finally obtain  the iteration relation,
\begin{eqnarray}
\langle a^{\dag}(\textbf{p}_{1}) a(\textbf{p}_{2}) \rangle_{N} = \nonumber \\
 \frac{Z_{N-1}(\sigma)}{Z_{N}(\sigma)} G_{1}^{*}(\textbf{p}_{2}, \textbf{p}_{1},\sigma)+ 
 \frac{Z_{N-1}(\sigma)}{Z_{N}(\sigma)}\int d^{3}k G_{1}^{*}(\textbf{p}_{2}, \textbf{k},\sigma)\langle a^{\dag}(\textbf{p}_{1}) a(\textbf{k}) \rangle_{N-1},
 \label{55}
\end{eqnarray}
which yields 
\begin{eqnarray}
\langle a^{\dag}(\textbf{p}_{1}) a(\textbf{p}_{2}) \rangle_{N} = \sum_{n=1}^{N}\frac{Z_{N-n}(\sigma)}{Z_{N}(\sigma)}  G_{n}^{*}(\textbf{p}_{2}, \textbf{p}_{1},\sigma).
 \label{56}
\end{eqnarray}

Now, let us derive an expression for the two-particle momentum 
spectra at a fixed particle number constraint,
\begin{eqnarray}
p^{0}_{1}p^{0}_{2} \frac{d^{6} N(N-1) }{d^{3}p_{1}d^{3}p_{2}} = p^{0}_{1}p^{0}_{2}  \langle a^{\dag}(\textbf{p}_{1}) a^{\dag}(\textbf{p}_{2}) a(\textbf{p}_{1}) a(\textbf{p}_{2}) \rangle_{N}.
\label{56.1} 
\end{eqnarray}
Using the cyclic invariance of the trace,  we obtain 
\begin{eqnarray}
\langle a^{\dag}(\textbf{p}_{1})a^{\dag}(\textbf{p}_{2}) a(\textbf{p}_{1})a(\textbf{p}_{2}) \rangle_{N}= \nonumber \\
 \frac{Z_{N-1}(\sigma)}{Z_{N}(\sigma)} \langle a^{\dag}(\textbf{p}_{1}) a(\textbf{p}_{1}) \rangle_{N-1} G_{1}^{*}(\textbf{p}_{2}, \textbf{p}_{2},\sigma) +  \frac{Z_{N-1}(\sigma)}{Z_{N}(\sigma)} \langle a^{\dag}(\textbf{p}_{2}) a(\textbf{p}_{1}) \rangle_{N-1} G_{1}^{*}(\textbf{p}_{2}, \textbf{p}_{1},\sigma)+ \nonumber \\
 \frac{Z_{N-1}(\sigma)}{Z_{N}(\sigma)}\int d^{3}k G_{1}^{*}(\textbf{p}_{2}, \textbf{k},\sigma)\langle a^{\dag}(\textbf{p}_{1})a^{\dag}(\textbf{p}_{2}) a(\textbf{p}_{1})a(\textbf{k}) \rangle_{N-1}.
 \label{57}
\end{eqnarray}
One can prove by induction that
\begin{eqnarray}
\langle a^{\dag}(\textbf{p}_{1})a^{\dag}(\textbf{p}_{2}) a(\textbf{p}_{1})a(\textbf{p}_{2}) \rangle_{N}= \nonumber \\
 \sum_{n=1}^{N}\frac{Z_{N-n}(\sigma)}{Z_{N}(\sigma)} \langle a^{\dag}(\textbf{p}_{1}) a(\textbf{p}_{1}) \rangle_{N-n} G_{n}^{*}(\textbf{p}_{2}, \textbf{p}_{2},\sigma) + \nonumber \\  \sum_{n=1}^{N}\frac{Z_{N-n}(\sigma)}{Z_{N}(\sigma)} \langle a^{\dag}(\textbf{p}_{2}) a(\textbf{p}_{1}) \rangle_{N-n} G_{n}^{*}(\textbf{p}_{2}, \textbf{p}_{1},\sigma).
 \label{58}
\end{eqnarray}
Combining  Eqs. (\ref{56}) and  (\ref{58}), we finally obtain 
\begin{eqnarray}
\langle a^{\dag}(\textbf{p}_{1})a^{\dag}(\textbf{p}_{2}) a(\textbf{p}_{1})a(\textbf{p}_{2}) \rangle_{N}= \nonumber \\
 \sum_{n=1}^{N-1} \sum_{s=1}^{N-n}  \frac{Z_{N-n-s}(\sigma)}{Z_{N}(\sigma)} \left( G_{n}^{*}(\textbf{p}_{2}, \textbf{p}_{2},\sigma) G_{s}^{*}(\textbf{p}_{1}, \textbf{p}_{1},\sigma) + G_{n}^{*}(\textbf{p}_{2}, \textbf{p}_{1},\sigma) G_{s}^{*}(\textbf{p}_{1}, \textbf{p}_{2},\sigma) \right).
 \label{59}
\end{eqnarray}
It is immediately apparent from the above expression that the computation of
$\langle a^{\dag}(\textbf{p}_{1})a^{\dag}(\textbf{p}_{2}) a(\textbf{p}_{1})a(\textbf{p}_{2}) \rangle_{N}$
involves summations over $n$ and $s$, and these summations fail to factorize.
Then, the question may arise as to  whether this expression is  invariant with respect to 
permutation  of particles, i.e., with respect to permutation $p_{1} \leftrightarrow p_{2}$. To address this 
question, let us note
that sums $ \sum_{n=1}^{N-1} \sum_{s=1}^{N-n} $ can be rewritten as $ \sum_{s=1}^{N-1} \sum_{n=1}^{N-s} $ . 
This means that Eq. (\ref{59}) is invariant  with respect to 
permutation $s \leftrightarrow n$ and, therefore, is invariant  with respect 
to permutation $p_{1} \leftrightarrow p_{2}$.

To evaluate  Eqs. (\ref{56}) and (\ref{59}), we need explicit  expressions for
$G_{n}^{*}(\textbf{p}_{2}, \textbf{p}_{1},\sigma)$
and the partition functions $Z_{n}(\sigma)$. The former has been evaluated in the previous section;
see Eq. (\ref{36.1}).
As for the latter, it  can be evaluated as follows. First, note that the definition of $\rho_{N} (\sigma)$
means that 
\begin{eqnarray}
\int d^{3}p \langle a^{\dag}(\textbf{p}) a(\textbf{p}) \rangle_{N} = N.
 \label{60}
\end{eqnarray}
Then, accounting for Eq. (\ref{56}), we get the recursive formula 
\begin{eqnarray}
nZ_{n} =  \sum_{s=1}^{n}Z_{n-s}\int d^{3}p G_{s}^{*}(\textbf{p}, \textbf{p},\sigma),
\label{61} 
\end{eqnarray}
where $Z_{0}=1$ by definition.

It is now a simple matter to write explicit expressions for the one- and two-particle momentum spectra. 
First, using  Eqs. (\ref{36.1}) and (\ref{61}), we get  the recurrence relation that can be easily implemented 
numerically, 
\begin{eqnarray}
nZ_{n} =  \sum_{s=1}^{n}Z_{n-s}\int d^{3}p \frac{1}{(2\pi)^{3}} \int_{\sigma}  \frac{d \sigma_{\mu}p^{\mu}}{p^{0}} e^{-s\beta_{\nu} (x)p^{\nu}} .
\label{62} 
\end{eqnarray}
Substituting Eq. (\ref{36.1}) into  Eq. (\ref{56}), we get
\begin{eqnarray}
\langle a^{\dag}(\textbf{p}_{1}) a(\textbf{p}_{2}) \rangle_{N} = \sum_{n=1}^{N}\frac{Z_{N-n}(\sigma)}{Z_{N}(\sigma)}  \frac{1}{(2\pi)^{3}\sqrt{p^{0}_{1}p^{0}_{2} }} \int_{\sigma}  d \sigma_{\mu}p^{\mu}e^{-i(p_{1}-p_{2})x} e^{-n\beta_{\nu} (x)p^{\nu}}.
 \label{62.1}
\end{eqnarray}
Consequently,   the one-particle momentum spectra at a fixed multiplicity 
constraint take the form 
 \begin{eqnarray}
p^{0} \frac{d^{3} N }{d^{3}p}= p^{0} \langle a^{\dag}(\textbf{p}) a(\textbf{p}) \rangle_{N} =  \int_{\sigma}  d \sigma_{\mu}p^{\mu} f^{\text{leq}}_{N}(x,p) ,
 \label{63}
 \end{eqnarray}
 where $f^{\text{leq}}_{N}(x,p)$ is the local-equilibrium canonical  distribution function at fixed $N$,
\begin{eqnarray}
f^{\text{leq}}_{N}(x,p) = \frac{1}{(2\pi)^{3}} \sum_{n=1}^{N}\frac{Z_{N-n}(\sigma)}{Z_{N}(\sigma)}  e^{-n\beta_{\nu} (x)p^{\nu}}.
 \label{64}
 \end{eqnarray}

It is worth noting that constant chemical potential
of the grand-canonical
ensemble, whose subensemble is the canonical fixed-$N$ ensemble, does not influence on particle momentum spectra 
and correlations calculated at fixed multiplicity. It follows  from the recurrence relation that
 $Z_{n}[\mu] = e^{\beta \mu n}Z_{n}[\mu =0] $, and therefore  $e^{\beta \mu n}$ is factored out from expressions for
 particle momentum spectra and correlations.
 
Comparing Eq. (\ref{64}) with Eq. (\ref{37.2}), one can conclude that the  selection of a fixed-$N$ 
subensemble of the corresponding
local-equilibrium grand-canonical ensemble results in nontrivial modifications of  distribution functions. 
In particular, the one-particle 
distribution function (\ref{64}) demonstrates multiplicity-dependent 
deviations in spacetime and momentum dependencies from the familiar 
local-equilibrium Bose ideal gas  distribution function; see Eq. (\ref{37.2}). 
In the next section, we compare particle momentum spectra and  correlations calculated  in the 
local-equilibrium grand-canonical
and canonical ensembles for
some simple but reliable for $p+p$ collisions model. 

\section{ Particle momentum spectra and correlations: comparison of the  ensembles }

It is instructive to compare our findings with the   treatment which is based on the grand-canonical ensembles (GCE)
where chemical potential, $\mu = \mbox{const}< m$, is taken such that mean 
particle number, $\langle N \rangle$, is equal to particle number, $N$, in 
the canonical ensembles (CE)  with a fixed multiplicity.  Such an approach is often used for the
sake of calculational convenience.
 Our simulations are performed for a
simple  hydro-inspired \cite{Bjorken} local-equilibrium  model of the longitudinally boost-invariant
 expanding system. 
 In this model, the longitudinal direction ($Z$ axis)
coincides with the beam direction,  and the
$4$-velocity is given by\footnote{ The initial  collision of the two approaching nuclei or nucleons 
results in  a rapid expansion, which at first proceeds in
the longitudinal direction. Here, for simplicity,  we do not take into account transverse expansion of a system.} 
\begin{eqnarray}
u^{\mu} = (t/\tau,0,0,z/\tau),
 \label{65}
 \end{eqnarray}
where $\tau=\sqrt{t^{2}-z^{2}}$ is the proper time. 
We assume that a local-equilibrium  state is defined at a
hypersurface with constant energy density in the comoving
coordinate system.  Then, $\beta (x) $ is constant on the corresponding
hypersurface, and such a three-dimensional hypersurface $\sigma$ is defined by a
constant $\tau$. It is convenient to parametrize $t$ and $z$ at this hypersurface 
as 
\begin{eqnarray}
t = \tau \cosh{\eta},
 \label{66} \\
 z=\tau \sinh{\eta},
 \label{67} 
 \end{eqnarray}
where $\eta$ is the longitudinal spatial rapidity, $\tanh {\eta}= v_{L}$, and $v_{L}=z/t$ is the 
longitudinal velocity.
This implies that 
\begin{eqnarray}
d\sigma_{\mu}=d\sigma n_{\mu} = d\sigma u_{\mu} =  \tau d\eta d^{2}\textbf{r}_{T}u_{\mu} ,
 \label{68} 
 \end{eqnarray}
where 
$\textbf{r}_{T} = (r_{x},r_{y})$ are the transverse Cartesian coordinates. 

This picture of an ultrarelativistic collision is, of course, not valid for large values of the spatial rapidity and
for large transverse distances. We assume that the system has
a finite transverse  size encoded in the limits of integration
over $r_{T}$: $0 < r_{T} < R_{T}$. As for the  longitudinal direction, the finiteness
of the system is provided by limits of integration over spatial rapidity $\eta$: $- \eta_{f}< \eta < \eta_{f}$.

The on-mass-shell particle $4$-momentum $p^{\mu}$ can be expressed
through the momentum rapidity $y$, $\tanh{y}=p_{z}/p_{0}$; transverse momentum $\textbf{p}_{T}$; and transverse mass
$m_{T}=\sqrt{\textbf{p}_{T}^{2}+m^{2}}$,
\begin{eqnarray}
p^{\mu} =  (m_{T}\cosh{y}, \textbf{p}_{T},m_{T}\sinh{y} ).
 \label{69} 
 \end{eqnarray}
Then, 
\begin{eqnarray}
p^{\mu}u_{\mu} =  m_{T}\cosh{(y - \eta)}.
 \label{70} 
 \end{eqnarray}

 For specificity and in order to compare the ensembles at  the extreme small-system limits,
 we utilize for numerical calculations
the set of parameters corresponding roughly to the values at
the system’s breakup in $p+p$ collisions at the LHC
energies. We take the particle’s mass as
of a charged pion, $m = 139.57$ MeV, and the temperature
$T= 150$ MeV (then the inverse temperature $\beta =1/T =1/150$ MeV$^{-1}$).
For $\tau$, we use $1.5$ fm$/c$. To account for finiteness of the system 
we assume that $R_{T} = 2 $  fm and $\eta_{f}=2$. 

One-particle momentum spectra in the canonical ensemble with fixed multiplicity constraint,  $p^{0}\frac{d^{3}N}{d^{3}p}$,  are calculated utilizing Eqs. (\ref{62}), (\ref{63}), and  (\ref{64}):
\begin{eqnarray}
p^{0}\frac{d^{3}N}{d^{3}p} = p^{0} \langle a^{\dag}(\textbf{p}) a(\textbf{p}) \rangle_{N}=\frac{1}{(2\pi)^{3}} \sum_{n=1}^{N}\frac{Z_{N-n}(\sigma)}{Z_{N}(\sigma)} \int_{\sigma}  d \sigma_{\mu}p^{\mu}  e^{-n\beta_{\nu} (x)p^{\nu}}. 
 \label{71} 
 \end{eqnarray}
 Employing  the  longitudinally 
boost invariant parametrization,  we get  
\begin{eqnarray}
\frac{d^{2}N}{2\pi m_{T}dm_{T}dy} = \frac{\pi R_{T}^{2}}{(2\pi)^{3}}  \sum_{n=1}^{N}\frac{Z_{N-n}(\sigma)}{Z_{N}(\sigma)}\Phi (n,m_{T},y),
 \label{72} 
 \end{eqnarray}
where $Z_{n}$ are defined by the recurrence relation  
\begin{eqnarray}
nZ_{n} = \frac{\pi R_{T}^{2}}{(2\pi)^{3}} \sum_{s=1}^{n}Z_{n-s}\int 2\pi m_{T}dm_{T}dy  \Phi (s,m_{T},y) ,
\label{73} 
\end{eqnarray}
and 
\begin{eqnarray}
\Phi (n,m_{T},y) =   \int_{\sigma} \frac{ \tau d\eta  m_{T}\cosh{(y - \eta)}}{ e^{n\beta m_{T}\cosh{(y - \eta)}}} .
\label{73.1} 
\end{eqnarray}

One-particle momentum spectra in the  grand-canonical ensemble with $\langle N \rangle = N$ are calculated utilizing Eqs. (\ref{37.1}) and  (\ref{37.2})  after substituting 
$\beta (x) p^{\nu}u_{\nu}(x) \rightarrow \beta (x) (p^{\nu}u_{\nu} (x) - \mu) $. Then, 
\begin{eqnarray}
p^{0}\frac{d^{3}\langle N \rangle }{d^{3}p} = p^{0} \langle a^{\dag}(\textbf{p}) a(\textbf{p}) \rangle =  \frac{1}{(2\pi)^{3}} \int_{\sigma}  d \sigma_{\mu}p^{\mu} \frac{1}{e^{\beta (x)(p^{\nu}u_{\nu} (x) - \mu)}-1}.
 \label{74} 
 \end{eqnarray}
For the considered model it implies that
 \begin{eqnarray}
 \frac{d^{2}\langle N \rangle}{2 \pi m_{T}dm_{T}dy} = \frac{1}{(2\pi)^{3}} \pi R_{T}^{2}\int_{\sigma}  \tau d\eta   \frac{m_{T}\cosh{(y - \eta)}}{e^{\beta (m_{T}\cosh{(y - \eta)} - \mu)}-1}. 
 \label{75} 
 \end{eqnarray}

We now turn to the two-particle momentum correlations. 
The two-particle momentum correlation function at fixed multiplicities is
defined as ratio of two-particle momentum spectrum to
one-particle ones and in the canonical ensemble with fixed particle number constraint can be evaluated as 
\begin{eqnarray}
C_{N}(\textbf{p}_{1},\textbf{p}_{2}) =G_{N}\frac{p^{0}_{1}p^{0}_{2} \frac{d^{6}N(N-1) }{d^{3}p_{1}d^{3}p_{2}}}{p^{0}_{1}\frac{d^{3} N  }{d^{3}p_{1}} p^{0}_{2}\frac{d^{3} N  }{d^{3}p_{2}} } = G_{N}\frac{ p^{0}_{1}p^{0}_{2} \langle a^{\dag}(\textbf{p}_{1})a^{\dag}(\textbf{p}_{2}) a(\textbf{p}_{1})a(\textbf{p}_{2}) \rangle_{N}}{p^{0}_{1}\langle a^{\dag}(\textbf{p}_{1})a(\textbf{p}_{1})  \rangle_{N} p^{0}_{2}\langle a^{\dag}(\textbf{p}_{2})a(\textbf{p}_{2})  \rangle_{N}  }, 
\label{76} 
\end{eqnarray}
where $\langle a^{\dag}(\textbf{p}_{1})a^{\dag}(\textbf{p}_{2}) a(\textbf{p}_{1})a(\textbf{p}_{2}) \rangle_{N}$ and $\langle a^{\dag}(\textbf{p}_{1})a(\textbf{p}_{1})  \rangle_{N}$ are defined in Eqs. (\ref{36.1}),
(\ref{59}), (\ref{63}), and  (\ref{64}). Here, $G_{N}$ is the normalization
constant. The latter is needed to normalize the
theoretical correlation function in accordance with normalization
that is applied by experimentalists: $C^{exp}_{N} \rightarrow 1$
for $|\textbf{p}_{1}-\textbf{p}_{2}| \rightarrow \infty$ and fixed $(\textbf{p}_{1}+\textbf{p}_{2})$.

It is convenient to evaluate the correlation function in terms of the
relative momentum $\textbf{q}=\textbf{p}_{2} - \textbf{p}_{1}$ and the pair momentum
$\textbf{k}=(\textbf{p}_{1} + \textbf{p}_{2})/2$.  The correlation function takes a particular 
simple form for pairs with vanishing longitudinal pair momentum  
$k_{z}=(p_{1z}+p_{2z})/2=0$ and with $\textbf{k}_{T}=\textbf{p}_{1T}= \textbf{p}_{2T}$, where 
$\textbf{k}_{T}$ is the 
pair momentum projected onto the transverse plane. Then, momentum 
rapidities of the particles in pairs are $y_{1}=-y_{2}$.  Explicitly, the longitudinal
projection ($\textbf{q}_{T}=\textbf{0}$)
of the correlation function, $C_{N}(\textbf{k}_{T},\textbf{q}_{L})$ (the subscript $L=$``long'' 
indicates the longitudinal direction), 
is given by 
\begin{eqnarray}
C_{N}(\textbf{k}_{T},q_{L}) = G_{N} (C_{N}^{(1)}(\textbf{k}_{T},q_{L}) +C_{N}^{(2)}(\textbf{k}_{T},q_{L})  ),
\label{77} 
\end{eqnarray}
where 
\begin{eqnarray}
C_{N}^{(1)}(\textbf{k}_{T},q_{L}) = \sum_{n=1}^{N-1} \sum_{s=1}^{N-n}  \frac{Z_{N-n-s}}{Z_{N}} \Phi (n,m_{T},y_{2})\Phi (s,m_{T},y_{1})  \times \nonumber \\   \left [ \sum_{n=1}^{N}\frac{Z_{N-n}(\sigma)}{Z_{N}(\sigma)} \Phi (n,m_{T},y_{1})\right ]^{-1} \left [ \sum_{n=1}^{N}\frac{Z_{N-n}(\sigma)}{Z_{N}(\sigma)} \Phi (n,m_{T},y_{2}) \right ]^{-1} , 
\label{78} 
\end{eqnarray}
and 
\begin{eqnarray}
C_{N}^{(2)} (\textbf{k}_{T},q_{L})=   \sum_{n=1}^{N-1} \sum_{s=1}^{N-n}  \frac{Z_{N-n-s}}{Z_{N}} \Psi (n,m_{T},y_{2},-q_{L})\Psi (s,m_{T},y_{1},q_{L})  \times \nonumber \\   \left [ \sum_{n=1}^{N}\frac{Z_{N-n}(\sigma)}{Z_{N}(\sigma)} \Phi (n,m_{T},y_{1})\right ]^{-1} \left [ \sum_{n=1}^{N}\frac{Z_{N-n}(\sigma)}{Z_{N}(\sigma)} \Phi (n,m_{T},y_{2}) \right ]^{-1}.
\label{79} 
\end{eqnarray}
Here
\begin{eqnarray}
\Psi (n,m_{T},y_{1},q_{L})=   \int_{\sigma}    \frac{\tau d\eta m_{T}\cosh{\eta}\cosh{y_{1}}e^{iq_{L}\tau \sinh{\eta}}}{e^{n\beta m_{T}\cosh{\eta}\cosh{y_{1}} }} .
\label{80} 
\end{eqnarray}
To completely  specify the two-boson correlation function (\ref{77}), one needs to estimate the  normalization
constant $G_{N}$. It can be realized by means of  the limit $ |q_{L} |\rightarrow
\infty$ at fixed $\textbf{k}_{T}$ in the corresponding expression. One
can readily see that proper normalization is reached if 
\begin{eqnarray}
G_{N}=
\frac{Z_{N}}{Z_{N-2}}\left(\frac{Z_{N-1}}{Z_{N}}\right
)^{2}.
 \label{81}
\end{eqnarray}

It is  of interest to estimate
the significance of the differences between the correlation functions 
calculated in the canonical and grand-canonical ensembles.
We define the correlation function in the  grand-canonical ensemble with $\langle N \rangle = N$ as 
\begin{eqnarray}
C(\textbf{p}_{1},\textbf{p}_{2}) = \frac{ p^{0}_{1}p^{0}_{2} \langle a^{\dag}(\textbf{p}_{1})a^{\dag}(\textbf{p}_{2}) a(\textbf{p}_{1})a(\textbf{p}_{2}) \rangle}{p^{0}_{1}\langle a^{\dag}(\textbf{p}_{1})a(\textbf{p}_{1})  \rangle p^{0}_{2}\langle a^{\dag}(\textbf{p}_{2})a(\textbf{p}_{2})  \rangle  } 
= \nonumber \\ 1+ \left|\int_{\sigma}   \frac{d \sigma_{\mu}p^{\mu}e^{-i(p_{1}-p_{2})x}}{e^{\beta (x)(p^{\nu}u_{\nu} (x) - \mu)}-1}\right |^{2} \left [ \int_{\sigma}   \frac{d \sigma_{\mu}p^{\mu}_{1}}{e^{\beta (x)(p^{\nu}_{1}u_{\nu} (x) - \mu)}-1}\right ]^{-1} \left [\int_{\sigma}   \frac{d \sigma_{\mu}p^{\mu}_{2}}{e^{\beta (x)(p^{\nu}_{2}u_{\nu} (x) - \mu)}-1} \right ]^{-1} 
\label{82} 
\end{eqnarray}
where $p^{\mu}=(p_{1}^{\mu}+p_{2}^{\mu})/2$.
Then, 
\begin{eqnarray}
C(\textbf{k}_{T},q_{L}) =   1+\left|\int_{\sigma}  \tau d\eta   \frac{m_{T}\cosh{\eta}\cosh{y_{1}}e^{-iq_{L}\tau \sinh{\eta}}}{e^{\beta (m_{T}\cosh{( \eta)} \cosh{y_{1}}- \mu)}-1}\right |^{2} \times \nonumber \\ \left [\int_{\sigma}  \tau d\eta   \frac{m_{T}\cosh{(y_{1} - \eta)}}{e^{\beta (m_{T}\cosh{(y_{1} - \eta)} - \mu)}-1}\right ]^{-1} \left [\int_{\sigma}  \tau d\eta   \frac{m_{T}\cosh{(y_{2} - \eta)}}{e^{\beta (m_{T}\cosh{(y_{2} - \eta)} - \mu)}-1} \right ]^{-1} .
\label{83} 
\end{eqnarray}

\begin{figure}[!ht]
\centering
\includegraphics[scale=0.7]{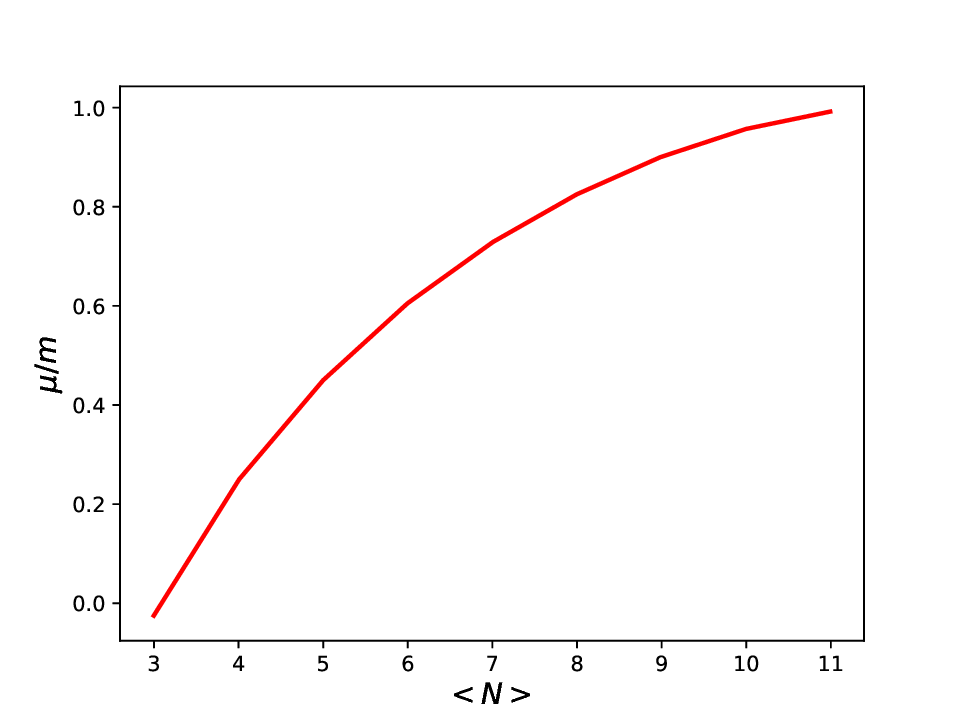}
\caption{The $\mu / m$ dependence on  $\langle N \rangle $. See the text for details. }
\label{fig:1}
\end{figure}

To compare the ensembles, we begin by calculating $\langle N \rangle$ as function of  $\mu /m$. 
The results are presented in Fig. \ref{fig:1}. One observes from this figure that $\mu$ is 
about  $m$ when $\langle N \rangle$ is near $11$. Because  we do not aim  to calculate here
the Bose-Einstein condensation in the grand-canonical and canonical ensembles, in what follows, we 
do not consider canonical ensembles with $N$ larger than $11$. 

\begin{figure}[!ht]
\centering
\includegraphics[scale=0.7]{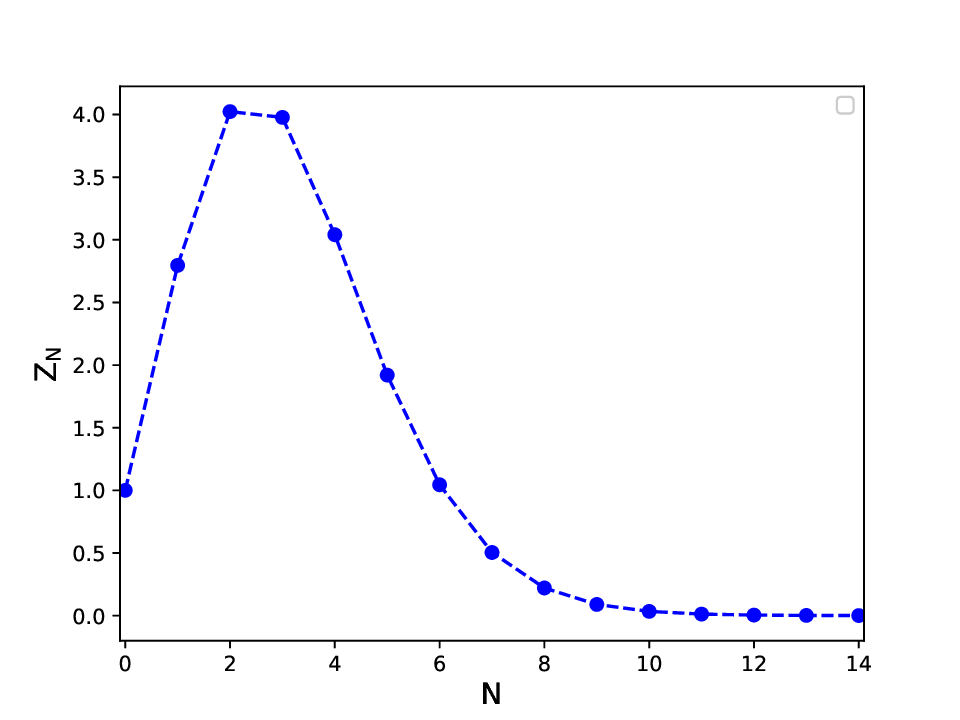}
\caption{The $Z_{N}$ dependence on $N$. See the text for details.}
\label{fig:2}
\end{figure}

Then, to calculate particle momentum spectra
and correlations in the canonical ensembles,  we need to evaluate $Z_{N}$ for various $N$. The 
results are plotted in Fig. \ref{fig:2}.

\begin{figure}[!ht]
\centering
\includegraphics[scale=0.7]{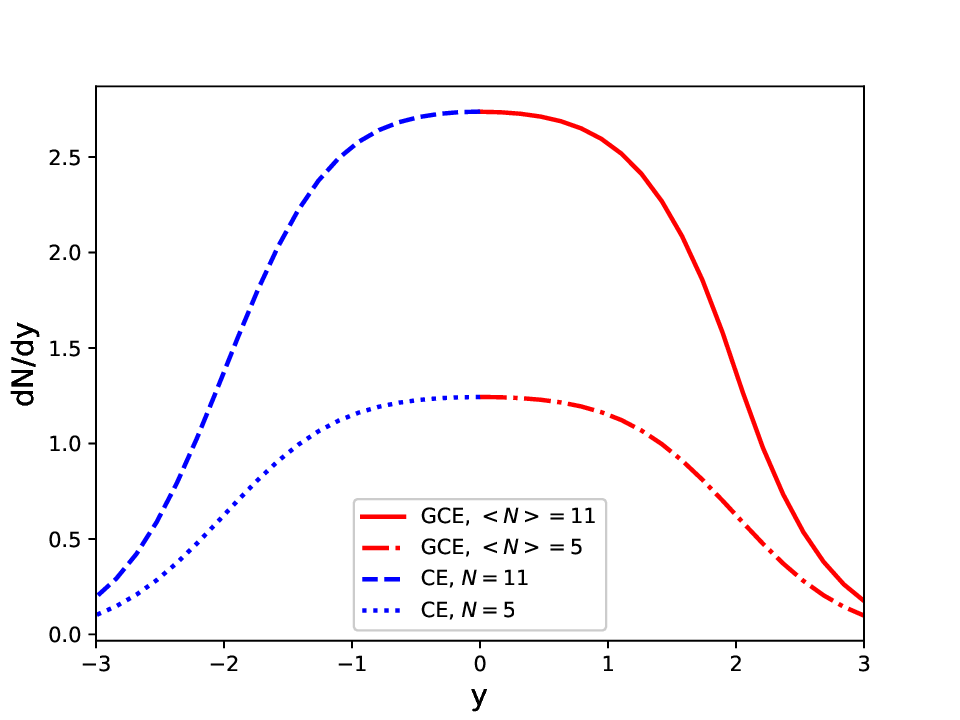}
\caption{The particle  rapidity densities for the canonical  (left) and grand-canonical ensembles (right). }
\label{fig:3}
\end{figure}

Now, we are ready to compare spectra and correlations calculated in the grand-canonical 
and canonical ensembles. First, we compare particle number rapidity  densities, $dN/dy$, 
for $\langle N \rangle =N$.  As illustrated by Fig. \ref{fig:3}, the grand-canonical
particle number rapidity  densities  are virtually indistinguishable from their 
canonical counterparts.

\begin{figure}[!ht]
\centering
\includegraphics[scale=0.7]{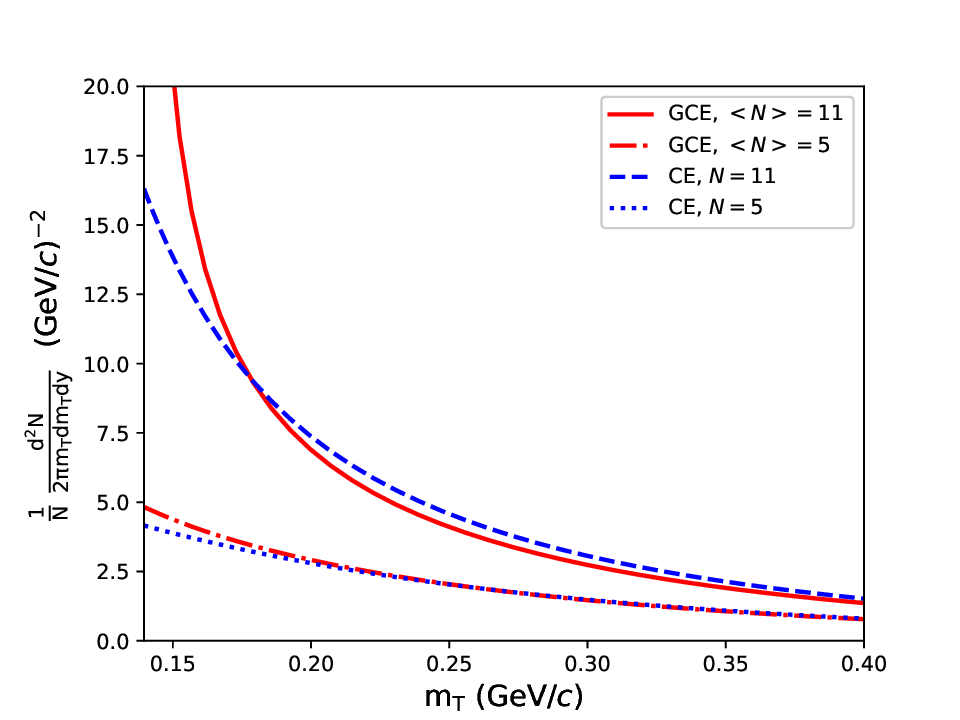}
\caption{The transverse momentum spectra  calculated in the canonical and grand-canonical ensembles with 
different $N = \langle N \rangle $. }
\label{fig:4}
\end{figure}

The transverse particle momentum spectra are compared in Fig. \ref{fig:4}. Aside from the low
transverse momenta region of the  spectra   with
$N=\langle N \rangle =11$, where the grand-canonical spectrum  is above the canonical one  due
to the Bose-Einstein enhancement ($\mu$ is approximately equal   
to $m$; see Fig. \ref{fig:1}), we see no significant differences.

\begin{figure}[!ht]
\centering
\includegraphics[scale=0.7]{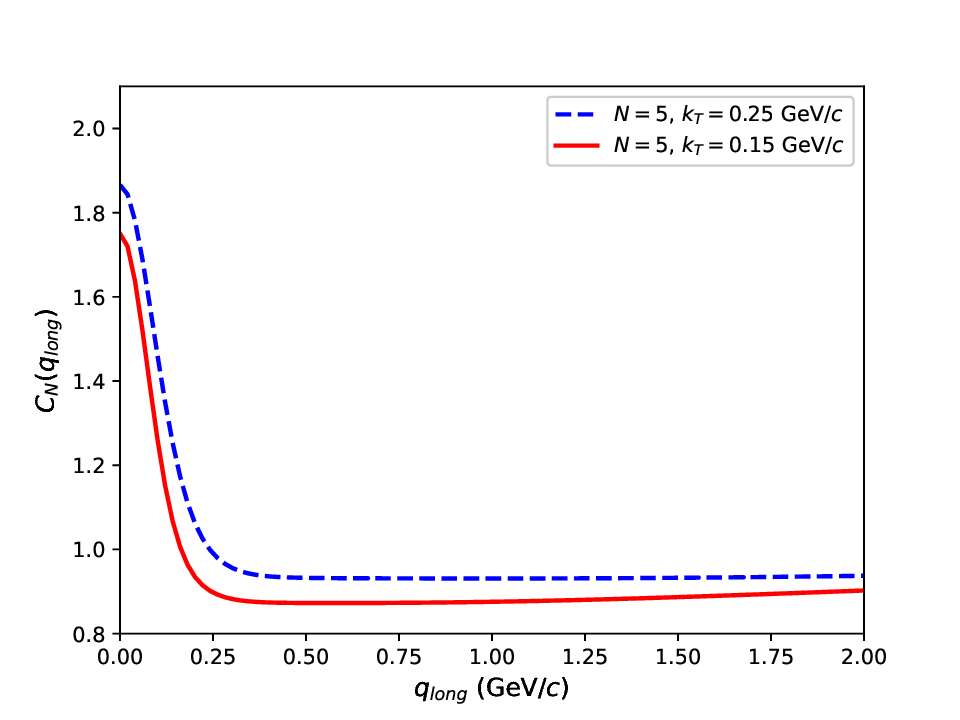}
\caption{The canonical correlation functions  for $N=5$ and several  different values of  $k_{T}$.  }
\label{fig:5}
\end{figure}

\begin{figure}[!ht]
\centering
\includegraphics[scale=0.7]{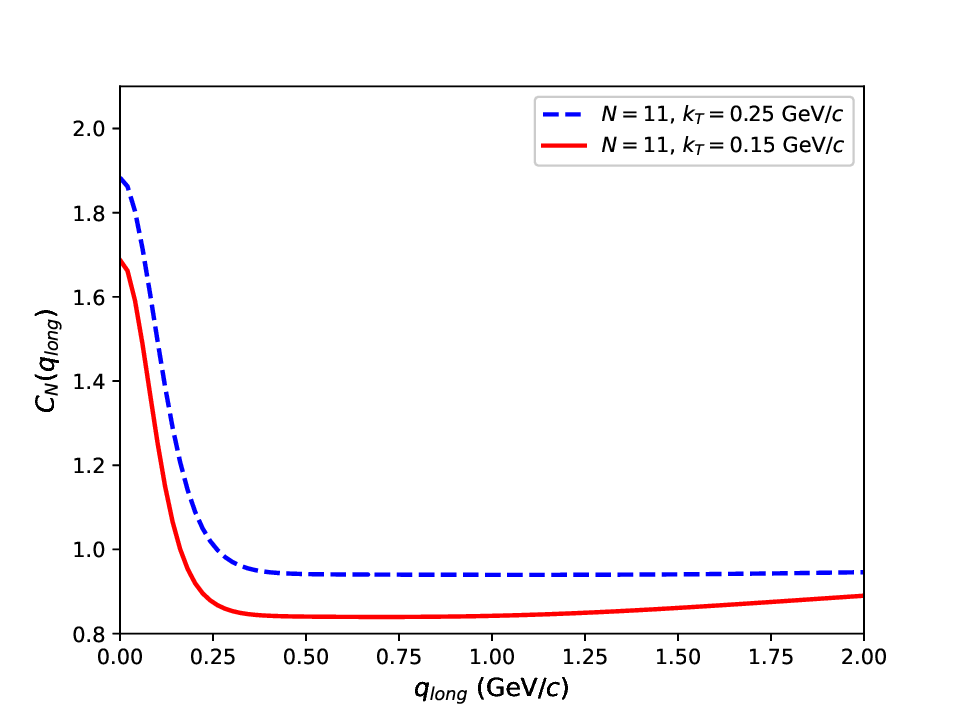}
\caption{The canonical correlation functions  for $N=11$ and several  different values of  $k_{T}$. }
\label{fig:6}
\end{figure}

Figures \ref{fig:5} and \ref{fig:6}  display  two-boson momentum correlation
functions $C_{N}(\textbf{k}_{T},q_{L})$ calculated in the canonical ensembles as a
function of the momentum difference. From these figures,  it is
evident that the intercepts of the canonical correlation functions, $C_{N}(\textbf{k}_{T},0)$, are not 
equal to $2$ and that the canonical correlation functions approach to $1$ from below when
$ |q_{L} |\rightarrow
\infty$.  It distinguishes two-boson correlation
functions in the canonical ensembles from the ones in
the corresponding grand-canonical ensembles where the
correlation functions (not shown here)  approach to $1$ from above  and the intercepts are  equal to $2$.

\begin{figure}[!ht]
\centering
\includegraphics[scale=0.7]{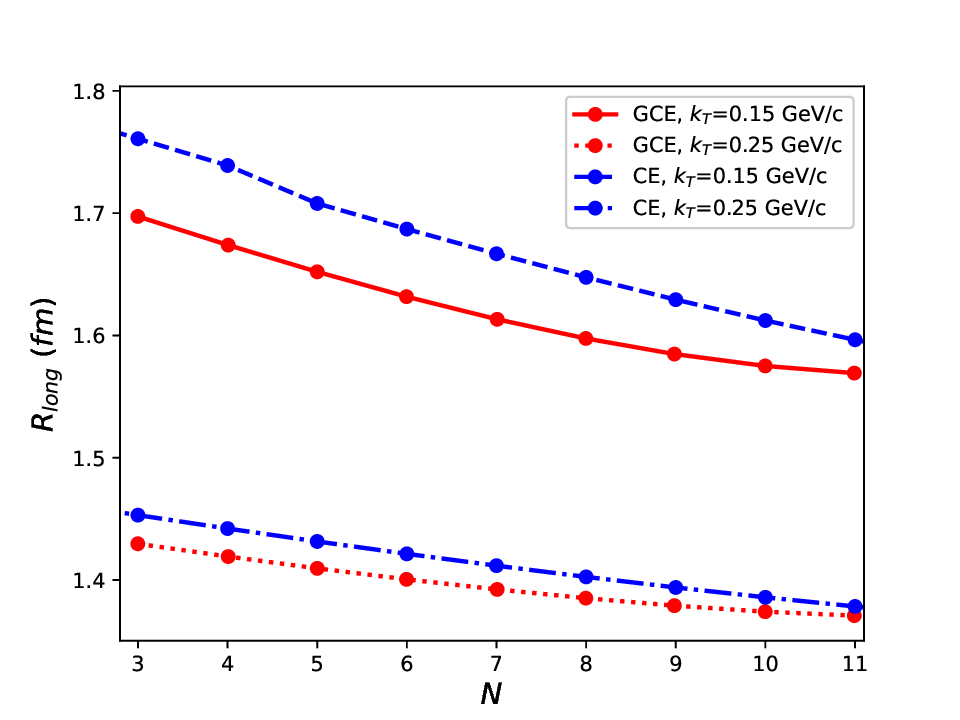}
\caption{The $R_{long}$ dependence on $N=\langle N \rangle$ in the canonical and grand-canonical ensembles
for several  different values of  $k_{T}$. See the text for details.}
\label{fig:7}
\end{figure}

\begin{figure}[!ht]
\centering
\includegraphics[scale=0.7]{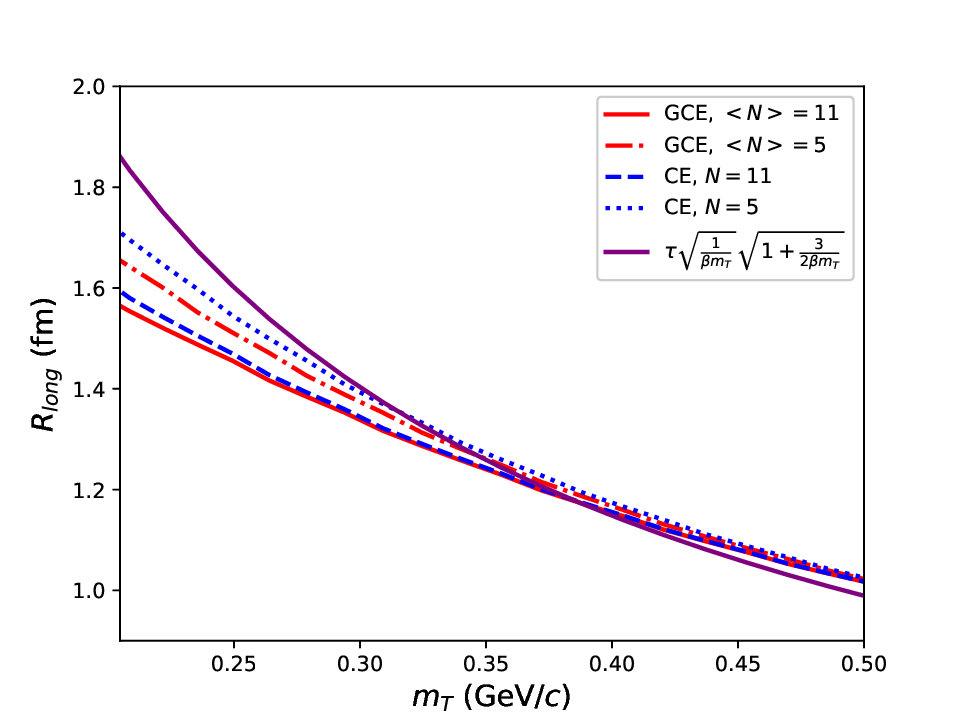}
\caption{The  $R_{long}$ dependence on $m_{T}$  in the canonical and grand-canonical ensembles  for several  different values of  $N=\langle N \rangle$ and the    $R_{long}$  calculated from the approximate analytic expression.   See the text for details. }
\label{fig:8}
\end{figure}

Notwithstanding the essential non-Gaussianity
of the canonical correlation functions, if the fitting procedure is
restricted to the correlation peak region, then the correlation
function is well fitted by the Gaussian expression 
\begin{eqnarray}
C_{N}(\textbf{k}_{T},q_{L}) = \frac{C_{N}(\textbf{k}_{T},0)}{2}\left ( 1+ e^{-q_{L}^{2}R_{long}^{2}(k_{T},N)} \right ) .
\label{84} 
\end{eqnarray}
It is instructive to compare canonical radius parameters extracted according to this expression
with the ones calculated in the grand-canonical ensembles for
$\langle N \rangle = N$,
\begin{eqnarray}
C(\textbf{k}_{T},q_{L}) =  1+ e^{-q_{L}^{2}R_{long}^{2}(k_{T},\langle N \rangle)} .
\label{85} 
\end{eqnarray}
For definiteness, for both ensembles we apply the fitting procedures in the $q_{L}$ range $0<q_{L}<q_{L}^{max}$,
where $q_{L}^{max}$ is such that $e^{-q_{L}^{2}R_{long}^{2}}=0.4$.
Our results are depicted in Figs. \ref{fig:7} and \ref{fig:8}. One can see that the canonical radius parameters slightly
decrease with $N$,  and the same trend, i.e., a decrease with $\langle N \rangle = N$, is also observed for 
the grand-canonical radius parameters that are slightly smaller  than the canonical ones. This decrease with 
$\langle N \rangle = N$ can be interpreted as increasing  deviations
from the Boltzmann approximation. Figure \ref{fig:8} shows $R_{long}$   as a
function on $k_{T} = |\textbf{k}_{T}|$ for several different values of $N$. One can see
that $R_{long}$ in both ensembles  is much smaller than the actual longitudinal size of the system
($\sim \tau \sinh{\eta_{f}}$)
and decreases when $k_{T}$ increases. Such a smallness of the correlation radius parameters  and  a decline
with increasing
pair momentum are  typical
for locally equilibrated expanding systems \cite{Sin-1}. 
In Fig. \ref{fig:8} we plot for comparison
the approximate analytical formula  for $R_{long}$, 
$R_{long} \approx \tau \sqrt{\frac{1}{\beta m_{T}}} \sqrt{\frac{K_{2}(\beta m_{T})}{K_{1}(\beta m_{T})}}\approx \tau \sqrt{\frac{1}{\beta m_{T}}}\sqrt{1+\frac{3}{2\beta m_{T}}}$ \cite{Ber}.
The latter approximate equality is obtained by means of the asymptotic (large argument)
expansion of the Macdonald functions. In the limit, $\beta m_{T} \gg 1$,  this reduces to the formula 
$R_{long} \approx \tau \sqrt{\frac{1}{\beta m_{T}}} $ \cite{Sin-2} (see also \cite{Sin-3}).
 All of the two figures reveal a consistent trend: 
if radius parameters are fitted in the region
of the correlation peak, then deviations of the canonical
radius parameters from their grand-canonical counterparts are rather small. It is also the case  for 
$N=\langle N \rangle =11$ and small $k_{T}$, because 
the effects of the Bose-Einstein enhancement ($\mu$ is approximately equal   
to $m$ at $\langle N \rangle =11$) are nearly canceled out in the ratio (\ref{83}).

\section{ Conclusions }

In this paper, we derived analytical expressions for  one- and two- particle momentum spectra  of a noninteracting
relativistic boson field in the canonical ensemble described by the local-equilibrium statistical operator
with a fixed particle 
number constraint. To see the effect of this constraint, we considered a corresponding grand-canonical state
and compared the one-particle spectra and two-particle Bose-Einstein correlation functions. The correspondence
was fixed by the condition that particle numbers, $N$, in the canonical states and mean particle
numbers, $\langle N \rangle$, in the grand-canonical states are the same. Then, applying hydrodynamically motivated
parametrization and  parameter values that
correspond roughly to the values at the system’s breakup
in $p+p$ collisions at the LHC energies, we compare our results with the  grand-canonical ensemble
where artificial chemical potential, $\mu = \mbox{const}< m$, is taken such that $\langle N \rangle =N$.
We have found that, calculated in both ensembles, one-particle 
momentum spectra are rather close to each other except for low
transverse momenta region of the spectra with $N=\langle N \rangle =11$,  where the grand-canonical spectrum 
is above the canonical one  due
to the Bose-Einstein enhancement ($\mu$ is approximately equal   
to $m$).
Then, we compared the two-particle Bose-Einstein momentum
correlations. We demonstrated  that there are  small quantitative but qualitative differences between 
the correlation radius parameters in both ensembles
if they are fitted  in the region of the correlation peak: the canonical radius parameters are slightly 
larger than the grand-canonical ones. Furthermore, we showed  that, in contrast 
to the predictions of the grand-canonical
ensemble, the intercepts of the canonical correlation functions are not equal to $2$ and 
depend on particle multiplicities  and momenta and that the  canonical 
correlation functions can be less than unity in some intermediate region of relative momentum of particles.
 Such  features should be taken into account when theoretical models are 
compared with  the multiplicity-dependent
measurements of the Bose-Einstein momentum correlations. As a final comment, we wish to note that
the  apparent independence of  correlation radius parameters on the particle number densities in high-multiplicity 
$p+p$ collisions at a fixed energy of the LHC   \cite{Atlas,CMS}   still remains unexplained, inviting further studies.

\begin{acknowledgments}
This work was supported by a grant from the Simons Foundation (Grant No. $1039151$, M.A. and S.A).
M.A. acknowledges support from the National Academy of Sciences of Ukraine priority project
``Properties of the matter at high energies and in galaxies during the epoch of the reionization
of the Universe'' (No. 0123U102248).
The research was carried out within the National Academy of 
Sciences of Ukraine Targeted 
Research Program ``Collaboration in advanced international projects on high-energy physics and nuclear physics'',
 Grant  No. $7/2023$ between the National Academy of 
Sciences of Ukraine and  Bogolyubov Institute for Theoretical 
Physics of the National Academy of Sciences of Ukraine. 
\end{acknowledgments}

\end{document}